\documentclass[11pt]{article}
\usepackage{amsmath,amssymb}
\usepackage[dvips]{epsfig}
\usepackage{graphicx}

{\hspace*{\fill}$\rule{.3\baselineskip}{.35\baselineskip}$\end{trivlist}}

\renewcommand{\phi}{\varphi}
\newcommand{\be}{\begin{eqnarray}}
\newcommand{\ee}{\end{eqnarray}}

\begin{document}

\title{\bf Integrable semi-discretization of the massive Thirring system in laboratory coordinates}

\author{Nalini Joshi$^{1}$ and Dmitry E. Pelinovsky$^{2}$  \\
{\small $^{1}$ School of Mathematics and Statistics F07, University of Sydney, NSW 2006, Australia} \\
{\small $^{2}$ Department of Mathematics and Statistics, McMaster University, Hamilton, Ontario, Canada, L8S 4K1} }

\date{\today}
\maketitle

\begin{abstract}
Several integrable semi-discretizations are known in the literature for the massive Thirring system
in characteristic coordinates. We present for the first time an integrable semi-discretization of the massive
Thirring system in laboratory coordinates. Our approach relies on the relation between the
continuous massive Thirring system and the Ablowitz--Ladik lattice. In particular, we
derive the Lax pair for the integrable semi-discretization of the massive Thirring system
by combining together the time evolution of the massive Thirring system in laboratory coordinates
and the B\"{a}cklund transformation for solutions to the Ablowitz--Ladik lattice.
\end{abstract}

\section{Introduction}

The purpose of this work is to find an integrable semi-discretization of the massive Thirring model (MTM)
in laboratory coordinates, which is a relativistically invariant nonlinear Dirac equation derived in general
relativity \cite{Thirring}. In the space of (1+1) dimensions, the MTM
can be written as a system of two semi-linear equations for $(u,v) \in \mathbb{C}^2$ in the normalized form:
\begin{equation}
\label{MTM}
\left\{ \begin{array}{l}
\displaystyle
i \left( \frac{\partial u}{\partial t} + \frac{\partial u}{\partial x} \right) + v  = |v|^2 u, \\
\displaystyle
i \left( \frac{\partial v}{\partial t} - \frac{\partial v}{\partial x} \right) + u = |u|^2 v. \end{array} \right.
\end{equation}
The standard Cauchy problem is posed in the time coordinates $t \in \mathbb{R}$
for the initial data $(u_0,v_0)$ extended in the spatial coordinate $x \in \mathbb{R}$.
Solutions of the Cauchy problem were studied recently in \cite{Candy1,Candy2,Huh11,Huh13,Huh15,SelTes,Zhang,ZhangZhao}.

Integrability of the MTM follows from the existence of the following Lax operators:
\begin{equation} \label{lablax1}
L(\lambda;u,v) = \frac{1}{2} (|u|^2-|v|^2) \sigma_3 +
\left[ \begin{matrix} 0 & \lambda v + \lambda^{-1} u \\ \lambda \bar{v} + \lambda^{-1} \bar{u} & 0 \end{matrix}\right]
+ \frac{1}{2}\left(\lambda^2 - \lambda^{-2} \right) \sigma_3,
\end{equation}
and
\begin{equation} \label{lablax2}
A(\lambda;u,v) = -\frac{1}{2}(|u|^2+|v|^2)\sigma_3 +
\left[\begin{matrix} 0 & \lambda v - \lambda^{-1} u \\ \lambda \bar{v} - \lambda^{-1} \bar{u} & 0 \end{matrix}\right]
+\frac{1}{2}\left(\lambda^2 + \lambda^{-2}\right)\sigma_3,
\end{equation}
where $\lambda$ is a spectral parameter, $\sigma_3 = {\rm diag}(1,-1)$ is the third Pauli spin matrix,
and $(\bar{u},\bar{v})$ stands for the complex conjugate of $(u,v)$.
The MTM system (\ref{MTM}) is the compatibility condition
\begin{equation}
\label{compatibility}
\frac{\partial^2 \vec{\phi}}{\partial x \partial t} = \frac{\partial^2 \vec{\phi}}{\partial t \partial x},
\end{equation}
for $\vec{\phi} \in \mathbb{C}^2$ satisfying the following Lax equations:
\begin{equation} \label{laxeq}
-2i \frac{\partial \vec{\phi}}{\partial x} = L(\lambda;u,v) \vec{\phi}\quad  \mbox{and}\quad -2i \frac{\partial \vec{\phi}}{\partial t} = A(\lambda;u,v) \vec{\phi}.
\end{equation}
Lax operators in the form (\ref{lablax1}) and (\ref{lablax2}) were introduced by Mikhailov \cite{Mikhailov}
and then studied in \cite{KN,KM}. More recently, the same Lax operators have been used for many purposes,
e.g. for the inverse scattering transform \cite{KMI,KW,PelSaal,Villarroel,W},
for spectral stability of solitary waves \cite{KL}, for orbital stability of Dirac solitons \cite{Yusuke1,Yusuke2},
and for construction of rogue waves \cite{Deg}.

Numerical simulations of nonlinear PDEs rely on spatial semi-discretizations obtained either with finite-difference
or spectral methods. Because the energy functional of the MTM system (\ref{MTM}) near the zero equilibrium $(u,v) = (0,0)$
is not bounded, either from above or below, the spatial discretizations of the MTM system (like in many other
massive Dirac models) suffer from spurious eigenvalues and other numerical instabilities \cite{BZ,Kevrekidis,Cuevas,Lakoba,Shao}.
An integrable semi-discretization would preserve the integrability scheme and model dynamics of nonlinear waves without
such spurious instabilities and other artifacts. This is why it is important
to derive an integrable semi-discretization of the MTM system (\ref{MTM}).

Since the pioneering works of Ablowitz and Ladik on equations related to the Ablowitz--Kaup--Newell--Segur (AKNS) spectral problem
\cite{AL,AL2}, it is well known that continuous nonlinear evolution equations integrable with the inverse scattering transform can be
semi-discretized in spatial coordinates or fully discretized in both spatial and temporal coordinates
in such a way as to preserve integrability \cite{Joshi-book}. Since the literature on this subject is vast,
we shall only restrict our attention to the relevant results on the massive Thirring model.

With the rotation of coordinates
\begin{equation}
\label{transformation}
\tau = \frac{1}{2} (t-x), \quad \xi = -\frac{1}{2} (x+t),
\end{equation}
the MTM system (\ref{MTM}) in laboratory coordinates $(x,t)$ can be rewritten in characteristic coordinates $(\xi,\tau)$:
\begin{equation}
\label{MTM-char}
\left\{ \begin{array}{l}
\displaystyle
-i \frac{\partial u}{\partial \xi} + v  = |v|^2 u, \\
\displaystyle
i \frac{\partial v}{\partial \tau} + u = |u|^2 v. \end{array} \right.
\end{equation}
The Cauchy problem in the time coordinate $\tau$ and the spatial coordinate $\xi$ for the system (\ref{MTM-char})
corresponds to the Goursat problem for the system (\ref{MTM}) and vice versa. Therefore, the spatial discretization
of the system (\ref{MTM-char}) in the spatial coordinate $\xi$ is not relevant for the Cauchy problem for
the system (\ref{MTM}). Unfortunately, it is the only integrable discretization available for the MTM up to now,
thanks to the relatively simple connection of the Lax operators for the system (\ref{MTM-char})
to the first negative flow of the Kaup--Newell (KN) spectral problem \cite{KN}.

The first result on the integrable discretizations of the MTM system in characteristic coordinates
goes back to the works of Nijhoff {\em et al.} in \cite{Nijhoff1,Nijhoff2}.
By using the B\"{a}cklund transformation for the continuous equations related to the KN spectral problem \cite{Nijhoff1},
integrable semi-discretizations in $\xi$ or full discretizations in $\xi$ and $\tau$ were obtained
for the MTM system (\ref{MTM-char}) and its equivalent formulations \cite{Nijhoff2}.
Since the relevant B\"{a}cklund transformation contains a square root singularity \cite{Nijhoff1}, the resulting
discretizations inherit a square root singularity \cite{Nijhoff2}, which may cause problems because
of ambiguity in the choice of square root branches and sign-indefinite expressions under the square root signs.
Unlike the continuous system (\ref{MTM-char}), the spatial discretizations constructed in \cite{Nijhoff2}
were not written in terms of the cubic nonlinear terms.

In a different direction, Tsuchida in \cite{Tsuchida} explored a gauge transformation of the KN spectral problem
to the AKNS spectral problem and constructed integrable semi-discretizations of nonlinear equations related
to the KN spectral problem. The semi-discretization constructed in \cite{Tsuchida} had cubic nonlinearity but
had a limitation of a different kind. The complex conjugate symmetry of the semi-discrete MTM system
was related to the lattice shift by half of the lattice spacing, where the variables were not defined.

In the latest work \cite{Tsuchida-preprint}, Tsuchida obtained another semi-discretization
of the MTM system (\ref{MTM-char}) by generalizing the Ablowitz--Ladik (AL) spectral
problem \cite{AL3} and B\"{a}cklund--Darboux transformations for nonlinear equations related to the AL spectral problem
\cite{Tsuchida2010,Vek,Zullo}. The new semi-discretization of the MTM system (\ref{MTM-char}) in \cite{Tsuchida-preprint} contains
the cubic nonlinearity and the complex conjugate reduction, which resemble those in the continuous system (\ref{MTM-char}).

How to transfer integrable semi-discretizations of the second-order equations
in characteristic coordinates (such as the sine--Gordon equation or the MTM system) to the integrable semi-discretizations
of these equations in laboratory coordinates has been considered to be an open problem for many years.
The main obstacle here is that the rotation of coordinates mixes positive and negative powers of the
spectral parameter $\lambda$ in the Lax operators related to the continuous case, hence the semi-discretization scheme needs to be revised.
At the same time, the temporal and spatial coordinates are already different in the semi-discrete case (one is continuous
and the other one is discrete) so that the rotation of coordinates produces a complicated difference-differential equation.
Since it is the Cauchy problem for the MTM system in laboratory coordinates that is used for most of physical applications
of the MTM system (\ref{MTM}), constructing a proper integrable semi-discretization of it becomes relevant and important.

The present work solves the problem stated above. We derive the semi-discrete
MTM system in laboratory coordinates together with its Lax pair by implementing the method of Tsuchida from
his recent work \cite{Tsuchida-preprint}. In particular, we combine together the time evolution of
the massive Thirring system in laboratory coordinates and the B\"{a}cklund transformation
for solutions to the Ablowitz--Ladik lattice. The main result is stated in Section \ref{sec-2}.
The proof of the main result appears in Section \ref{sec-3}. The conclusion is given in Section \ref{sec-4}.

\section{Main result}
\label{sec-2}

We start with the gauge-modified Lax operators for the MTM system (\ref{MTM}) derived by Barashenkov and Getmanov in \cite{BG87}:
\begin{equation} \label{lablax1-new}
\mathcal{L}(\lambda;u,v) = \left[\begin{matrix} \lambda^2 - |v|^2 & \lambda v + \lambda^{-1} u \\
\lambda \bar{v} + \lambda^{-1} \bar{u} & \lambda^{-2} - |u|^2 \end{matrix}\right],
\end{equation}
and
\begin{equation} \label{lablax2-new}
\mathcal{A}(\lambda;u,v) = \left[\begin{matrix} \lambda^2 - |v|^2 & \lambda v - \lambda^{-1} u \\
\lambda \bar{v} - \lambda^{-1} \bar{u} & -\lambda^{-2} + |u|^2 \end{matrix}\right].
\end{equation}
The gauge-modified Lax formulation (\ref{lablax1-new})--(\ref{lablax2-new}) differs from
the classical (zero-trace) Lax formulation (\ref{lablax1})--(\ref{lablax2}) only
in the diagonal terms. The MTM system (\ref{MTM}) still arises as the compatibility condition (\ref{compatibility}),
where $\vec{\phi} \in \mathbb{C}^2$ satisfies the Lax equations
\begin{equation} \label{laxeq-new}
-2i \frac{\partial \vec{\phi}}{\partial x} = \mathcal{L}(\lambda;u,v) \vec{\phi}\quad  \mbox{and}\quad
-2i \frac{\partial \vec{\phi}}{\partial t} = \mathcal{A}(\lambda;u,v) \vec{\phi},
\end{equation}
which are related to the new operators $\mathcal{L}$ and $\mathcal{A}$ in (\ref{lablax1-new})--(\ref{lablax2-new}).

The main result of this work is a derivation of the following spatial discretization
of the MTM system (\ref{MTM}) suitable for the Cauchy problem in the laboratory coordinates:
\begin{equation}
\label{MTM-discrete}
\left\{ \begin{array}{l}
\displaystyle
4 i \frac{d U_n}{dt} + Q_{n+1} + Q_n  + \frac{2i}{h} (R_{n+1}-R_n) + U_n^2 (\bar{R}_n + \bar{R}_{n+1}) \\
\qquad - U_n (|Q_{n+1}|^2 + |Q_n|^2 + |R_{n+1}|^2 + |R_n|^2) - \frac{ih}{2} U_n^2 (\bar{Q}_{n+1}-\bar{Q}_n)  = 0, \\
\displaystyle
-\frac{2i}{h} (Q_{n+1}-Q_n) + 2 U_n - |U_n|^2 (Q_{n+1} + Q_n) = 0, \\
\displaystyle
R_{n+1} + R_n - 2 U_n + \frac{ih}{2} |U_n|^2 (R_{n+1} - R_n) = 0, \end{array} \right.
\end{equation}
where $h$ is the lattice spacing parameter and $n$ is the discrete lattice variable. In the limit $h \to 0$
the slowly varying solutions between the lattice nodes can be represented by
$$
U_n(t) = U(x=hn,t), \quad R_n(t) = R(x = hn,t), \quad Q_n(t) = Q(x=nh,t),
$$
where the field variables satisfy the continuum limit of the system (\ref{MTM-discrete}) given by
\begin{equation}
\label{MTM-continuous}
\left\{ \begin{array}{l}
\displaystyle
2 i \frac{\partial U}{\partial t} + i \frac{\partial R}{\partial x} + Q + U^2 \bar{R} - U (|Q|^2 + |R|^2) = 0, \\
\displaystyle
-i \frac{\partial Q}{\partial x} + U - |U|^2 Q = 0, \\
\displaystyle
R - U = 0. \end{array} \right.
\end{equation}
The system (\ref{MTM-continuous}) in variables $U(x,t) = u(x,t-x)$ and $Q(x,t) = v(x,t-x)$ yields
the MTM system (\ref{MTM}). Therefore, the system (\ref{MTM-discrete}) is a proper
integrable spatial semi-discretization of the continuous MTM system in laboratory coordinates.

Note that the last two difference equations of the system (\ref{MTM-discrete}) are uncoupled
between the components $\{ R_n \}_{n \in \mathbb{Z}}$ and $\{ Q_n \}_{n \in \mathbb{Z}}$
and, moreover, they are linear with respect to $\{ R_n \}_{n \in \mathbb{Z}}$ and $\{ Q_n \}_{n \in \mathbb{Z}}$.
If the sequence $\{ U_n \}_{n \in \mathbb{Z}}$ decays to zero as $|n| \to \infty$ sufficiently fast,
then one can express $R_n$ and $Q_n$ in an explicit form by
\begin{equation}
\label{expression-R-n}
R_n = 2 \sum_{k=-\infty}^{n-1} (-1)^{n-k-1} U_k \frac{\prod_{m=k+1}^{n-1} (1 - i h |U_m|^2/2)}{\prod_{m=k}^{n-1} (1 + i h |U_m|^2/2)},
\end{equation}
and
\begin{equation}
\label{expression-Q-n}
Q_n = -ih \sum_{k=-\infty}^{n-1} U_k \frac{\prod_{m=k+1}^{n-1} (1 + i h |U_m|^2/2)}{\prod_{m=k}^{n-1} (1 - i h |U_m|^2/2)}.
\end{equation}
The time evolution of $U_n$ is obtained by solving the first differential equation of the system
(\ref{MTM-discrete}). Although the representations (\ref{expression-R-n}) and (\ref{expression-Q-n})
express $\{ R_n \}_{n \in \mathbb{Z}}$ and $\{ Q_n \}_{n \in \mathbb{Z}}$ in terms of $\{ U_n \}_{n \in \mathbb{Z}}$,
it may be easier computationally to solve the last two difference equations of the system (\ref{MTM-discrete})
instantaneously at every time $t \in \mathbb{R}$.

As follows from our construction, the semi-discrete MTM system (\ref{MTM-discrete}) is the compatibility condition
\begin{equation}
\label{Lax-discrete}
-2i\frac{d}{dt} N_n = P_{n+1} N_n - N_n P_n,
\end{equation}
for $\vec{\phi} \in \mathbb{C}^2$ on $n \in \mathbb{Z}$ and $t \in \mathbb{R}$ satisfying the following Lax equations:
\begin{equation}
\label{Lax-N-P}
\vec{\phi}_{n+1} = N_n \vec{\phi}_n, \quad -2i \frac{d\vec{\phi}_n}{dt} = P_n \vec{\phi}_n,
\end{equation}
associated with the Lax operators
\begin{equation}
\label{Lax-N}
N_n := \left[ \begin{array}{cc}
\lambda + 2i h^{-1} \lambda^{-1} \frac{1 + i h |U_n|^2/2}{1 - i h |U_n|^2/2}  & \frac{2 U_n}{1 - i h |U_n|^2/2}  \\
\frac{2 \bar{U}_n}{1 - i h |U_n|^2/2}  & -\lambda \frac{1 + i h |U_n|^2/2}{1 - i h |U_n|^2/2}  + 2i h^{-1} \lambda^{-1} \end{array} \right]
\end{equation}
and
\begin{equation}
\label{Lax-P}
P_n := \left[\begin{matrix} \lambda^2 - |R_n|^2 & \lambda R_n - \lambda^{-1} Q_n  \\
\lambda \bar{R}_n - \lambda^{-1} \bar{Q}_n & -\lambda^{-2} + |Q_n|^2 \end{matrix}\right].
\end{equation}
The trivial zero solution satisfies the semi-discrete MTM system (\ref{MTM-discrete})
and reduces the Lax equations (\ref{Lax-N-P}) to uncoupled equations
for components of $\vec{\phi}$ which are readily solvable. Non-trivial solutions
of the semi-discrete MTM system (\ref{MTM-discrete}) will be constructed in future work.

\section{Proof of the main result}
\label{sec-3}

Here we derive the semi-discrete MTM system (\ref{MTM-discrete}) by using the method of Tsuchida \cite{Tsuchida-preprint} with suitable modifications.
Since the work \cite{Tsuchida-preprint} relies on the MTM system in characteristic coordinates (\ref{MTM-char})
and the Ablowitz--Ladik lattice, we will incorporate these details in our subsequent analysis.
The proof is broken into several subsections.

\subsection{MTM system in characteristic coordinates}

The MTM system (\ref{MTM-char}) is the compatibility condition
\begin{equation}
\frac{\partial^2 \vec{\phi}}{\partial \xi \partial \tau} = \frac{\partial^2 \vec{\phi}}{\partial \tau \partial \xi}.
\end{equation}
for $\vec{\phi} \in \mathbb{C}^2$ satisfying the following Lax equations:
\begin{equation} \label{Lax-equation}
i \frac{\partial \vec{\phi}}{\partial \xi} = \left[ \begin{array}{cc} \lambda^2 - |v|^2 & \lambda v \\
 \lambda \bar{v} & 0 \end{array} \right] \vec{\phi} \quad
 \mbox{\rm and} \quad
i \frac{\partial \vec{\phi}}{\partial \tau} = \left[ \begin{array}{cc} 0 & \lambda^{-1} u \\
 \lambda^{-1} \bar{u} & \lambda^{-2} - |u|^2 \end{array} \right] \vec{\phi}.
\end{equation}
Consequently, by using the transformation (\ref{transformation}) in the Lax equations
(\ref{Lax-equation}), we obtain the Lax equations (\ref{laxeq}) with $L$ and $A$ given by
(\ref{lablax1-new}) and (\ref{lablax2-new}).

The Lax equations (\ref{Lax-equation}) with triangular matrices
are different from the classical Lax equations
with zero-trace matrices \cite{KM}:
\begin{equation} \label{Lax-equation-classical}
i \frac{\partial \vec{\phi}}{\partial \xi} = \left[ \begin{array}{cc} \frac{1}{2} (\lambda^2 - |v|^2) & \lambda v \\
 \lambda \bar{v} & -\frac{1}{2} (\lambda^2 - |v|^2) \end{array} \right] \vec{\phi},
\quad
i \frac{\partial \vec{\phi}}{\partial \tau} = \left[ \begin{array}{cc} -\frac{1}{2} (\lambda^{-2} - |u|^2) & \lambda^{-1} u \\
 \lambda^{-1} \bar{u} & \frac{1}{2} (\lambda^{-2} - |u|^2) \end{array} \right] \vec{\phi}.
\end{equation}
where the $\xi$-dependent problem is referred to as Kaup--Newell spectral problem \cite{KN}
and the $\tau$-dependent problem is the first negative flow of the Kaup--Newell hierarchy.
As shown in \cite{BG87}, the two formulations are gauge-equivalent by using the conservation law
\begin{equation}
\label{MTM-conservation}
\frac{\partial |v|^2}{\partial \tau} = \frac{\partial |u|^2}{\partial \xi},
\end{equation}
which holds for the MTM system in characteristic coordinates (\ref{MTM-char}).
Adding $\frac{1}{2} (\lambda^2 - |v|^2) I_{2 \times 2}$ to the $\xi$-dependent problem
and $\frac{1}{2} (\lambda^{-2} - |u|^2) I_{2 \times 2}$ to the $\tau$-dependent problem
in the classical Lax equations (\ref{Lax-equation-classical}) with zero-trace matrices, where
$I_{2 \times 2}$ is the $2 \times 2$ identity matrix, yields
the Lax equations (\ref{Lax-equation}) with triangular matrices.
The representations (\ref{Lax-equation}) and (\ref{Lax-equation-classical}) 
are equivalent because the MTM system
(\ref{MTM-char}) is invariant under the gauge transformation of the Lax operators \cite{BG87}.

\subsection{Relation to the Ablowitz--Ladik lattice}

The gauge-modified Lax formulation (\ref{Lax-equation}) of the MTM system (\ref{MTM-char})
is related to the following Ablowitz--Ladik (AL) lattice:
\begin{equation} \label{AL-lattice}
\left\{ \begin{array}{l}
\displaystyle
\frac{dQ_m}{dt} = a (1 - Q_m R_m) (Q_{m+1} - Q_{m-1}) + ib (1-Q_m R_m) (Q_{m+1} + Q_{m-1}), \\
\displaystyle
\frac{dR_m}{dt} = a (1 - Q_m R_m) (R_{m+1} - R_{m-1}) - i b (1-Q_m R_m) (R_{m+1} + R_{m-1}),\end{array} \right.
 \quad m \in \mathbb{Z},
\end{equation}
where $a,b \in \mathbb{R}$ are parameters of the model.
In order to show how the AL lattice (\ref{AL-lattice}) is related to the MTM system (\ref{MTM-char}),
we write the Lax equations for the AL lattice:
\begin{equation}
\label{Lax-equation-AL}
\vec{\phi}_{m+1} = \left[ \begin{array}{cc} \lambda & Q_m \\
R_m & \lambda^{-1} \end{array} \right] \vec{\phi}_m,
\end{equation}
and
\begin{equation}
\label{Lax-equation-AL-time}
\frac{d \vec{\phi}_m}{\partial t} =
(a+ib) \left[ \begin{array}{cc} \lambda^2 - Q_m R_{m-1} & \lambda Q_m \\
 \lambda R_{m-1} & 0 \end{array} \right] \vec{\phi}_m + (a-ib)
\left[ \begin{array}{cc} 0 & \lambda^{-1} Q_{m-1} \\
 \lambda^{-1} R_m & \lambda^{-2} - Q_{m-1} R_m \end{array} \right] \vec{\phi}_m.
\end{equation}
The compatibility condition for the system (\ref{Lax-equation-AL})--(\ref{Lax-equation-AL-time})
yields the AL lattice (\ref{AL-lattice}). With the correspondence
$$
Q_{m-1} = u, \quad Q_{m} = v, \quad R_{m-1} = \bar{v}, \quad R_{m} = \bar{u},
$$
the two matrix operators in the linear combination of the time evolution equation
(\ref{Lax-equation-AL-time}) can be used
in the Lax equations (\ref{Lax-equation}). In this context, the index $m$ can be dropped
and the variables $(u,v)$ satisfy the MTM system (\ref{MTM-char})
from commutativity of the Lax equations (\ref{Lax-equation}).

\subsection{B\"{a}cklund--Darboux transformation for the AL lattice}

It is known \cite{Vek,Zullo} that a new solution $\{\tilde{Q}_m,\tilde{R}_m\}_{m \in \mathbb{Z}}$
of the AL lattice (\ref{AL-lattice}) can be obtained from another solution $\{Q_m,R_m\}_{m \in \mathbb{Z}}$
of the same lattice by the B\"{a}cklund--Darboux transformation. The B\"{a}cklund--Darboux transformation
also relates the eigenvectors $\vec{\phi}_m$ and $\tilde{\vec{\phi}}_m$
satisfying the Lax equations (\ref{Lax-equation-AL}) and (\ref{Lax-equation-AL-time})
for the same spectral parameter $\lambda$.
The relation between $\vec{\phi}_m$ and $\tilde{\vec{\phi}}_m$
can be written in the form used in \cite{Tsuchida-preprint}:
\begin{equation}
\label{BT}
\tilde{\vec{\phi}}_m =  \left[ \begin{array}{cc} \alpha \lambda + \delta \lambda^{-1} & 0 \\
0 & \gamma \lambda + \beta \lambda^{-1} \end{array} \right] \vec{\phi}_m
+ (\alpha \beta - \gamma \delta) \left[ \begin{array}{cc} \gamma \lambda & U_m \\
V_m & \delta \lambda^{-1} \end{array} \right]^{-1} \vec{\phi}_m,
\end{equation}
where $(\alpha,\beta,\gamma,\delta)$ are arbitrary parameters such that $\alpha \beta - \gamma \delta \neq 0$
and $\{ U_m,V_m\}_{m \in \mathbb{Z}}$ are some potentials. For the purpose of the B\"{a}cklund--Darboux transformation
for the AL lattice, the potentials $\{ U_m,V_m\}_{m \in \mathbb{Z}}$ are expressed in terms of eigenfunctions
satisfying the spectral problem (\ref{Lax-equation-AL}) at a fixed value of the spectral parameter $\lambda$.
However, for the purpose of the integrable semi-discretization of the MTM system, we specify constraints on
$\{ U_m,V_m\}_{m \in \mathbb{Z}}$ from the commutativity condition below.

\subsection{Derivation of the Lax equations (\ref{Lax-N-P}) with (\ref{Lax-N}) and (\ref{Lax-P})}

Let us drop the index $m$ and forget about the AL lattice (\ref{AL-lattice}) and the spectral problem (\ref{Lax-equation-AL}).
If the B\"{a}cklund--Darboux transformation (\ref{BT})
is iterated in new index $n$, we can introduce the new spectral problem
\begin{equation}
\label{BT-Lax}
\vec{\phi}_{n+1} = N_n \vec{\phi}_n
\end{equation}
with
$$
N_n := \left[ \begin{array}{cc} \alpha \lambda + \delta \lambda^{-1} & 0 \\
0 & \gamma \lambda + \beta \lambda^{-1} \end{array} \right]
+ (\alpha \beta - \gamma \delta) \left[ \begin{array}{cc} \gamma \lambda & U_n \\
V_n & \delta \lambda^{-1} \end{array} \right]^{-1},
$$
The spectral problem (\ref{BT-Lax}) is coupled with the time evolution problem
\begin{equation} \label{Time-Lax}
-2i \frac{d \vec{\phi}_n}{\partial t} = P_n \vec{\phi}_n,
\end{equation}
with
$$
P_n := \left[\begin{matrix} \lambda^2 - |R_n|^2 & \lambda R_n - \lambda^{-1} Q_n  \\
\lambda \bar{R}_n - \lambda^{-1} \bar{Q}_n & -\lambda^{-2} + |Q_n|^2 \end{matrix}\right].
$$
The time evolution problem (\ref{Time-Lax}) is obtained from the time evolution problem 
in (\ref{lablax2-new}) and (\ref{laxeq-new}) for the MTM system (\ref{MTM}).
In the Lax equations (\ref{BT-Lax}) and (\ref{Time-Lax}), we have unknown
potentials $\{ U_n, V_n, R_n, Q_n\}_{n \in \mathbb{Z}}$ and arbitrary parameters
$(\alpha,\beta,\gamma,\delta)$ such that $\alpha \beta - \gamma \delta \neq 0$.

In what follows, we consider the compatibility condition (\ref{Lax-discrete})
and obtain the constraints on these unknown potentials.
Since the spectral parameter $\lambda$ is independent of $t$, we obtain
$$
\frac{d}{dt} N_n = - (\alpha \beta - \gamma \delta)
\left[ \begin{array}{cc} \gamma \lambda & U_n \\
V_n & \delta \lambda^{-1} \end{array} \right]^{-1}
\left[ \begin{array}{cc} 0 & \frac{d U_n}{dt} \\
\frac{d V_n}{dt} & 0 \end{array} \right]
\left[ \begin{array}{cc} \gamma \lambda & U_n \\
V_n & \delta \lambda^{-1} \end{array} \right]^{-1}.
$$
For $N_n$ in $P_{n+1} N_n$, we can use the equivalent representation
$$
N_n = \left[ \begin{array}{cc} \alpha \lambda (\gamma \lambda + \beta \lambda^{-1}) & U_n (\alpha \lambda + \delta \lambda^{-1}) \\
V_n (\gamma \lambda + \beta \lambda^{-1}) & \beta \lambda^{-1} (\alpha \lambda + \delta \lambda^{-1}) \end{array} \right]
 \left[ \begin{array}{cc} \gamma \lambda & U_n \\
V_n & \delta \lambda^{-1} \end{array} \right]^{-1}.
$$
For $N_n$ in $N_n P_n$, we can use the equivalent representation
$$
N_n =  \left[ \begin{array}{cc} \gamma \lambda & U_n \\
V_n & \delta \lambda^{-1} \end{array} \right]^{-1}
\left[ \begin{array}{cc} \alpha \lambda (\gamma \lambda + \beta \lambda^{-1}) & U_n (\gamma \lambda + \beta \lambda^{-1}) \\
V_n (\alpha \lambda + \delta \lambda^{-1}) & \beta \lambda^{-1} (\alpha \lambda + \delta \lambda^{-1}) \end{array} \right].
$$

\subsubsection{(1,1) and (2,2) entries}

Substituting the equivalent expressions for $N_n$ to the Lax equation (\ref{Lax-discrete}) yields the following two constraints
arising at different powers of $\lambda$ in the $(1,1)$ entries:
\begin{eqnarray}
\label{constraint-11a}
\alpha \gamma (|R_{n+1}|^2 - |R_n|^2) + \gamma (U_n \bar{R}_n - V_n R_{n+1}) + \alpha (V_n R_n - U_n \bar{R}_{n+1}) & = & 0, \\
\label{constraint-11b}
U_n V_n (|Q_{n}|^2 - |Q_{n+1}|^2) + \alpha (U_n \bar{Q}_{n+1} - V_n Q_n) + \gamma (V_n Q_{n+1} - U_n \bar{Q}_{n}) & = & 0
\end{eqnarray}
and the following two constraints arising at different powers of $\lambda$ in the $(2,2)$ entries:
\begin{eqnarray}
\label{constraint-22a}
U_n V_n (|R_{n+1}|^2 - |R_n|^2) + \beta (U_n \bar{R}_n - V_n R_{n+1}) + \delta (V_n R_n - U_n \bar{R}_{n+1}) & = & 0, \\
\label{constraint-22b}
\beta \delta (|Q_{n}|^2 - |Q_{n+1}|^2) + \delta (U_n \bar{Q}_{n+1} - V_n Q_n) + \beta (V_n Q_{n+1} - U_n \bar{Q}_{n}) & = & 0.
\end{eqnarray}

We claim that the constraints (\ref{constraint-11a})--(\ref{constraint-22b}) are equivalent to the following constraints:
\begin{eqnarray}
\label{constraint1}
\gamma (\alpha \beta - U_n V_n) R_{n+1} - \alpha (\delta \gamma - U_n V_n) R_n & = & (\alpha \beta - \delta \gamma) U_n, \\
\label{constraint2}
\alpha (\gamma \delta - U_n V_n) \bar{R}_{n+1} - \gamma (\alpha \beta - U_n V_n) \bar{R}_n & = & -(\alpha \beta - \delta \gamma) V_n,
\end{eqnarray}
and
\begin{eqnarray}
\label{constraint3}
\beta (\gamma \delta - U_n V_n) Q_{n+1} - \delta (\alpha \beta - U_n V_n) Q_n & = & -(\alpha \beta - \delta \gamma) U_n, \\
\label{constraint4}
\delta (\alpha \beta - U_n V_n) \bar{Q}_{n+1} - \beta (\gamma \delta - U_n V_n) \bar{Q}_n & = & (\alpha \beta - \delta \gamma) V_n.
\end{eqnarray}
To show the equivalence, we use linear combinations of (\ref{constraint-11a}) and (\ref{constraint-22a}) and obtain
\begin{eqnarray*}
\alpha (\gamma \delta - U_n V_n) (|R_{n+1}|^2 - |R_n|^2) - (\alpha \beta - \gamma \delta) (U_n \bar{R}_n - V_n R_{n+1}) & = & 0, \\
\gamma (\alpha \beta - U_n V_n) (|R_{n+1}|^2 - |R_n|^2) + (\alpha \beta - \gamma \delta) (V_n R_n - U_n \bar{R}_{n+1}) & = & 0.
\end{eqnarray*}
These relations are regrouped as follows:
\begin{eqnarray*}
R_{n+1} \left[ \alpha (\gamma \delta - U_n V_n) \bar{R}_{n+1} + (\alpha \beta - \gamma \delta) V_n \right] -
\bar{R}_n \left[ \alpha (\gamma \delta - U_n V_n) R_n + (\alpha \beta - \gamma \delta) U_n \right] & = & 0, \\
\bar{R}_{n+1} \left[ \gamma (\alpha \beta - U_n V_n) R_{n+1} - (\alpha \beta - \gamma \delta) U_n \right] -
R_n \left[ \gamma (\alpha \beta - U_n V_n) \bar{R}_n - (\alpha \beta - \gamma \delta) V_n \right] & = & 0,
\end{eqnarray*}
where each constraint is satisfied if and only if constraints (\ref{constraint1}) and (\ref{constraint2}) hold.
The equivalence of constraints (\ref{constraint-11b}) and (\ref{constraint-22b}) to constraints (\ref{constraint3}) and (\ref{constraint4})
is established by similar operations.

\subsubsection{(1,2) entries}

At different powers of $\lambda$ in the $(1,2)$ entries of the Lax equation (\ref{Lax-discrete}), we obtain two constraints
\begin{eqnarray}
\label{constraint-12a}
\alpha \gamma U_n (|R_{n+1}|^2 - |R_n|^2) + (\alpha \beta - \gamma \delta) U_n +
\alpha \gamma (\delta R_n - \beta R_{n+1}) + U_n^2 (\gamma \bar{R}_n - \alpha \bar{R}_{n+1}) & = & 0, \\
\label{constraint-12b}
\beta \delta U_n (|Q_{n}|^2 - |Q_{n+1}|^2) + (\alpha \beta - \gamma \delta) U_n +
\beta \delta (\gamma Q_{n+1} - \alpha Q_{n}) + U_n^2 (\delta \bar{Q}_{n+1} - \beta \bar{Q}_{n}) & = & 0,
\end{eqnarray}
and the evolution equation
\begin{eqnarray}
\nonumber
2i (\alpha \beta - \gamma \delta) \frac{d U_n}{dt} + \alpha \gamma (\beta Q_{n+1} - \delta Q_n)
+ \beta \delta (\alpha R_n - \gamma R_{n+1}) + U_n^2 (\alpha \bar{Q}_{n+1} - \gamma \bar{Q}_n) \\
\label{evolution-1}
+ U_n^2 (\beta \bar{R}_n - \delta \bar{R}_{n+1})
+ U_n (\gamma \delta |Q_n|^2 - \alpha \beta |Q_{n+1}|^2) + U_n (\gamma \delta |R_{n+1}|^2 - \alpha \beta |R_n|^2) = 0.
\end{eqnarray}
We claim that the two constraints (\ref{constraint-12a}) and (\ref{constraint-12b}) are redundant in view of the
constraints (\ref{constraint1})--(\ref{constraint4}). Indeed, substituting
$$
\alpha \gamma (\beta R_{n+1} -\delta R_n) - (\alpha \beta - \delta \gamma) U_n = U_n V_n (\gamma R_{n+1} - \alpha R_n)
$$
from (\ref{constraint1}) into (\ref{constraint-12a}) and dividing the result by $U_n$, we obtain
$$
\alpha \gamma (|R_{n+1}|^2 - |R_n|^2) - V_n(\gamma R_{n+1} - \alpha R_n)  + U_n (\gamma \bar{R}_n - \alpha \bar{R}_{n+1}) = 0,
$$
which is nothing but (\ref{constraint-11a}). Similar transformations apply to (\ref{constraint-12b}) with (\ref{constraint3})
to end up at (\ref{constraint-22b}). Hence the constraints (\ref{constraint-12a}) and (\ref{constraint-12b}) are satisfied if
the constraints (\ref{constraint1})--(\ref{constraint4}) hold.

\subsubsection{(2,1) entries}

Similarly, at different powers of $\lambda$ in the $(2,1)$ entries of the Lax equation (\ref{Lax-discrete}), we obtain two constraints
\begin{eqnarray}
\label{constraint-21a}
\alpha \gamma V_n (|R_{n+1}|^2 - |R_n|^2) - (\alpha \beta - \gamma \delta) V_n +
\alpha \gamma (\beta \bar{R}_n - \delta \bar{R}_{n+1}) + V_n^2 (\alpha R_n - \gamma R_{n+1}) & = & 0, \\
\label{constraint-21b}
\beta \delta V_n (|Q_{n}|^2 - |Q_{n+1}|^2) - (\alpha \beta - \gamma \delta) V_n +
\beta \delta (\alpha \bar{Q}_{n+1} - \gamma \bar{Q}_{n}) + V_n^2 (\beta Q_{n+1} - \delta Q_{n}) & = & 0,
\end{eqnarray}
and the evolution equation
\begin{eqnarray}
\nonumber
2i (\alpha \beta - \gamma \delta) \frac{d V_n}{dt} + \alpha \gamma (\delta \bar{Q}_{n+1} - \beta \bar{Q}_n)
+ \beta \delta (\gamma \bar{R}_n - \alpha \bar{R}_{n+1}) + V_n^2 (\gamma Q_{n+1} - \alpha Q_n) \\
\label{evolution-2}
+ V_n^2 (\delta R_n - \beta R_{n+1})
+ V_n (\alpha \beta |Q_n|^2 - \gamma \delta |Q_{n+1}|^2) + V_n (\alpha \beta  |R_{n+1}|^2 - \gamma \delta |R_n|^2) = 0.
\end{eqnarray}
The two constraints (\ref{constraint-21a}) and (\ref{constraint-21b}) are once again redundant in view of the
constraints (\ref{constraint1})--(\ref{constraint4}). The proof of this is achieved by transformations
similar to the computations of $(1,2)$ entries.

Summarizing it up, we have shown that the Lax equation (\ref{Lax-discrete})
is satisfied under the four constraints (\ref{constraint1})--(\ref{constraint4})
and the two evolution equations (\ref{evolution-1}) and (\ref{evolution-2}).

\subsection{Choice of parameters}

The constraints (\ref{constraint3})--(\ref{constraint4}) and the time evolution equations (\ref{evolution-1})
and (\ref{evolution-2})
with $R_n = 0$ are obtained in \cite{Tsuchida-preprint}, where the B\"{a}cklund--Darboux transformation (\ref{BT-Lax})
is coupled with the time flow given by the negative powers of $\lambda$. Similarly,
the constraints (\ref{constraint1})--(\ref{constraint2}) and the time evolution equations (\ref{evolution-1})
and (\ref{evolution-2})
with $Q_n = 0$ are obtained when the B\"{a}cklund--Darboux transformation (\ref{BT-Lax})
is coupled with the time flow given by the positive powers of $\lambda$.
All four constraints (\ref{constraint1})--(\ref{constraint4})
and the full system of time evolution equations (\ref{evolution-1}) and (\ref{evolution-2}) arise
in the full time flow (\ref{Time-Lax}).

The complex-conjugate symmetry $V_n = \bar{U}_n$ between the constraints (\ref{constraint1}) and (\ref{constraint3}) on one side
and the constraints (\ref{constraint2}) and (\ref{constraint4}) on the other side as well as between the evolution equations
(\ref{evolution-1}) and (\ref{evolution-2}) is preserved if $\alpha = \bar{\gamma}$ and $\beta = \bar{\delta}$.
Without loss of generality, we normalize $\alpha = \gamma = 1$ and introduce the parameter $h \in \mathbb{R}$
by $\beta = -\delta = 2i/h$. Then, equations (\ref{constraint1}), (\ref{constraint3}), and (\ref{evolution-1})
divided by $2i/h$ give respectively the third, second, and first equations of the system (\ref{MTM-discrete}).
Thus, integrability of the system (\ref{MTM-discrete}) is verified from the Lax equations (\ref{Lax-N-P})
with the Lax operators given by (\ref{Lax-N}) and (\ref{Lax-P}).

\section{Conclusion}
\label{sec-4}

We have derived an integrable semi-discretization of the MTM system in laboratory coordinates (\ref{MTM}).
Integrability of the semi-discrete MTM system (\ref{MTM-discrete}) with the Lax operators (\ref{Lax-N}) and (\ref{Lax-P})
is a starting point for derivation of conserved quantities, multi-soliton solutions, and other useful facts of the semi-discrete MTM
in laboratory coordinates. It is also a starting point for numerical simulations of the integrable semi-discretizations
of the continuous MTM system (\ref{MTM}).

We conclude by mentioning some relevant works on the semi-discretizations of other integrable
nonlinear evolution equations.

By the semi-discretization of the Hirota bilinear method,
one can obtain an integrable semi-discretization of many continuous nonlinear equations as done
in \cite{ChenChen} for the coupled Yajima--Oikawa system. One can verify \cite{Pel-personal} that
this technique applied to the
Chen--Lee--Liu system yields the same semi-discretization as the one obtained by Tsuchida \cite{Tsuchida-preprint}
from coupling the B\"{a}cklund transformation for the AL lattice and the time evolution of the Chen--Lee--Liu system.

From a different point of view, Lax operators for the AL-type lattices
with quadratic and higher-order polynomial dependence were considered by Vakhnenko
(see the recent review in \cite{Vakhnenko} and earlier references therein).
This approach brings many interesting semi-discretizations of coupled nonlinear Schr\"{o}dinger equations,
but these equations are not related to the semi-discretizations of the massive Thirring model \cite{Bronsard}.

We conclude that the present contribution
contains the only integrable semi-discretization of the MTM system in the laboratory coordinates available in the present time.

\vspace{0.5cm}

{\bf Acknowledgement.} D.E.P. thanks P.G. Kevrekidis for suggesting a search for integrable semi-discretization of the MTM system back in 2015,
as well as Th. Ioannidou and S.A. Bronsard for collaboration on the early (unsuccessful) stages of the project in 2016 and 2017
respectively. Critical advices from T. Tsuchida
helped the authors to achieve the goal. This work was completed while D.E.P. visited Department of Mathematics at
University of Sydney during January-June 2018. The research of N.J. was supported by an Australian Laureate Fellowship
\# FL 120100094 and grant \# DP160101728.

\end{document}